\documentclass[prb,preprint,showpacs]{revtex4}

\usepackage{graphicx}
\usepackage{dcolumn}
\usepackage{amssymb}
\usepackage{bm}
\usepackage{epsfig}
\usepackage{psfrag}

\begin{document}

\title{Collapse of Double-Walled Carbon Nanotube Bundles under Hydrostatic Pressure}

\author{Vikram Gadagkar,$^{1}$ Prabal K. Maiti,$^{1,*}$ Yves Lansac,$^{2}$ A. Jagota,$^{3}$ and A. K. Sood$^{1,}$}
\altaffiliation{To whom correspondence should be addressed:
asood@physics.iisc.ernet.in or maiti@physics.iisc.ernet.in}
\affiliation{$^{1}$Department of Physics, Indian Institute of
Science, Bangalore 560012, INDIA} \affiliation{$^{2}$LEMA, UMR
6157 CNRS-CEA, Universit\'e Fran\c{c}ois Rabelais, 37200 Tours,
FRANCE} \affiliation{$^{3}$Department of Chemical Engineering,
Lehigh University, Bethlehem, Pennsylvania 18015, USA}

\date{\today}

\begin{abstract}
We use classical molecular dynamics simulations to study the
collapse of single (SWNT) and double-walled (DWNT) carbon nanotube
bundles under hydrostatic pressure. The collapse pressure
($p_{c}$) varies as $1/R^{3}$, where $R$ is the SWNT radius or the
DWNT effective radius. The bundles show $\sim$ 30 \% hysteresis
and the hexagonally close packed lattice is completely restored on
decompression. The $p_{c}$ of DWNT is found to be close to the sum
of its values for the inner and the outer tubes considered
separately as SWNT, demonstrating that the inner tube supports the
outer tube and that the effective bending stiffness of DWNT,
$D_{DWNT}{\sim}2D_{SWNT}$. We use an elastica formulation to
derive the scaling and the collapse behavior of DWNT and
multi-walled carbon nanotubes.
\end{abstract}

\maketitle

\section{Introduction}

Since their discovery, carbon nanotubes have been subject to
intense theoretical and experimental investigations due to their
fascinating structural, electronic, and mechanical properties
\cite{Saito1998}. Carbon nanotubes are showing great promise in
such diverse fields as nanoelectronics, actuators, sensors
\cite{Ghosh2003}, nanofluidics, hydrogen storage, and
high-strength materials. The mechanical properties of carbon
nanotubes depend on the number of coaxial graphitic rings that go
into their making. Significant advances have been made in the
understanding of single (SWNT) and multi-walled (MWNT) carbon
nanotubes. Double-walled carbon nanotubes (DWNT) have been
observed and synthesized \cite{Smith1998,Bandow2001} more
recently. Being the simplest of the MWNT, they are ideal systems
to study the evolution of various properties from the single to
the multi-walled regime.

High-pressure Raman experiments on SWNT bundles
\cite{Venkateswaran1999,Peters2000,Sandler2003} point to a
structural phase transition at $\sim2$ GPa. The current
understanding is that the initially circular nanotube cross
section is distorted to an oval shape under pressure. High
pressure X-ray diffraction studies also indicate a phase
transition \cite{Sharma2001} from the ambient triangular lattice
symmetry, which reappears under decompression. Molecular dynamics
simulations  suggest that SWNT bundles \cite{Elliot2004,
Zhang2004} as well as isolated tubes \cite{Capaz2004,Sun2004}
collapse under hydrostatic pressure and that the collapse pressure
varies as an inverse power law of the tube radius. More recently,
several workers
\cite{Pfeiffer2003,Arvanitidis2005,Puech2004,Puech2004a,Venkateswaran2004}
have used Raman spectroscopy to study bundles of DWNT under
hydrostatic pressure. They conclude that the environment inside
the outer tube is highly defect free and unperturbed, that the
outer tube acts as a protective shield for the inner tube and that
the inner tube provides structural support to the outer tube.

In this paper, we describe a set of molecular dynamics simulations
performed to investigate the behavior of DWNT under pressure, focusing
on the response of the inner and the outer tubes. These results
are contrasted with similar MD simulations on bundles of SWNT. Observed
results are interpreted within the framework of the elastica
theory.

\section{Simulation Methodology}

We have used DREIDING \cite{Mayo1990}, a standard generic
macromolecular force field, in all our molecular dynamics (MD)
simulations. Table \ref{Table1} lists the force field parameters
used to calculate intra and inter molecular interactions. Elliott
et al. \cite{Elliot2004} have successfully used this force field
to study the collapse of SWNT bundles under hydrostatic pressure.
Our simulations have been performed using ModulaSim
\cite{Modulasim}, a modular and general purpose molecular modeling
package. The ensemble used was one of constant particle number,
pressure, and temperature (NPT). The temperature (300 K) and the
applied hydrostatic pressure were maintained using the Berendsen
thermostat and barostat \cite{Berendsen1984}. The simulation cell
consisted of 16 independent SWNT or DWNT arranged in a hexagonally
close packed 4 $\times$ 4 bundle, with periodic boundary
conditions and pressure applied along all three mutually
perpendicular directions. The tubes were ten unit cells long (2.3
nm). It has been found \cite{Elliot2004} that nine independent
tubes, ten unit cells long, are sufficient to avoid finite size
effects. The MD simulations were carried out on four SWNT, (5,5),
(10,10), (15,15), and (20,20) and four DWNT, (5,5)@(10,10),
(7,7)@(12,12), (10,10)@(15,15), and (15,15)@(20,20) bundles using
the standard velocity Verlet algorithm to integrate the equations
of motion. The gap between the inner and the outer tubes is $\sim$
3.4 $\AA$, close to the inter-layer gap in graphite. The bundles
were initially equilibrated at atmospheric pressure and
subsequently subjected to step-wise monotonically increasing
hydrostatic pressure increments, allowing the unit cell volume to
equilibrate for at least 10 ps at each step. The simulation time
step was 1 fs. Information about the structural transition was
obtained by measuring the unit cell volume after equilibration at
each hydrostatic pressure step.

\section{Results and Discussion}

All the SWNT and DWNT equilibrated at atmospheric pressure have
nearly circular cross sections, as shown in Figs. 1(a) and (c) for
a SWNT and a DWNT bundle, respectively. At atmospheric pressure,
$r_{min}/r_{max}\geq{0.93}$ where $r_{min}$ and $r_{max}$ are the
smallest and the largest distances from the center to the
circumference of the tube cross section. To study the structural
transition, we plot the reduced volume ($V/V_{0}$), where $V_{0}$
is the unit cell volume at atmospheric pressure, for the various
SWNT bundles as a function of pressure as shown in Fig. 2 (a). It
is clear that each of the four SWNT bundles undergoes a
spontaneous structural transition at a critical pressure
($p_{c}$), which decreases as the radius of the tubes increases,
in agreement with previously published results
\cite{Elliot2004,Zhang2004}. Unless otherwise specified, $p_{c}$
refers to the structural change pressure on the loading curve. On
plotting the reduced volume versus pressure for the DWNT bundles,
as shown in Fig. \ref{fig:2} (b), we once again observe clear
structural transitions at well-defined critical pressures. Up to
the critical pressure, the tube cross sections remain nearly
circular with slight deformations from the circular shape. When
the applied hydrostatic pressure exceeds $p_{c}$, the tube cross
sections assume an elliptical shape. Further increase in pressure
results in a dumbbell shape as shown in Figs. 1(b) and (d). The
loading and unloading curves show a $\sim$ 30\% hysteresis in all
the bundles studied. The hysteresis is calculated as
$100\%\times[p_{c}^{loading}-p_{c}^{unloading}]/p_{c}^{loading}$.
The hexagonally close packed lattice is completely restored in all
SWNT and DWNT bundles on decompression. A closer look at the
critical pressures of the DWNT bundles in Fig. \ref{fig:2} (b)
reveals several interesting features. First, we notice that the
$p_{c}$ of a DWNT bundle is greater than the $p_{c}$ of an SWNT
bundle of the outer tubes alone. For example, the $p_{c}$ of the
(10,10)@(15,15) DWNT is 4.1 GPa, a value higher as compared to the
$p_{c}$ of (15,15) SWNT (0.9 GPa). This shows that the inner tube
supports the outer tube under hydrostatic pressure. Second, the
$p_{c}$ of the DWNT bundle is even higher than the $p_{c}$ of an
SWNT bundle of the inner tubes alone. For the (10,10)@(15,15)
tubes, 4.1 GPa ($p_{c}$ of the DWNT bundle) is higher than 3.2 GPa
($p_{c}$ of (10,10) SWNT bundle). Having demonstrated that the
$p_{c}$ of a DWNT bundle is higher than the $p_{c}$ of both the
inner and the outer tubes, we now ask whether one can predict the
$p_{c}$ of a DWNT bundle with the knowledge of the $p_{c}$ of the
inner and the outer tubes. From Fig. \ref{fig:2}, we see that the
$p_{c}$ of a DWNT bundle is close to the sum of the $p_{c}$ of the
inner and the outer tubes. In our example, 4.1 GPa ($p_{c}$ of the
(10,10)@(15,15) DWNT bundle) is equal to 3.2 GPa ($p_{c}$ of
(10,10) SWNT bundle) plus 0.9 GPa ($p_{c}$ of (15,15) SWNT
bundle). In section IV, we derive an analytical result that
demonstrates this behavior.

Fig. \ref{fig:3} (a) shows the collapse pressure of the SWNT
bundles versus the tube radius along with a $1/R^{3}$ fit
\cite{Capaz2004,Sun2004}. If we now define an effective radius of
a MWNT with $n$ walls, as
$\frac{1}{R_{eff}^{3}}=\frac{1}{n}\sum_{i=1}^{n}\frac{1}{R_{i}^{3}}$,
the collapse pressure of the DWNT bundles is found to follow a
$1/R_{eff}^{3}$ dependence as seen from Fig. \ref{fig:3} (b).  Let
$D$ be the bending modulus of the graphene sheet so that the
energy per unit surface area associated with curvature $k$ is
given by \cite{yakobson1,Pantano2004,Tang_2cyl,Tang2005}
$u_{e}=\frac{D}{2}k^{2}$. The value of $D_{SWNT}$ has been
estimated from a plot of the single point energies per unit
surface area of seven isolated SWNT as a function of $1/R^{2}$
(Fig. \ref{fig:4} (a)), which gives the value of $D_{SWNT}$ to be
2.90 eV. The mean curvature for a bundle was calculated by
averaging its local value at each atom of each tube (see
Appendix). Using the values of curvature and $D$ (from Fig.
\ref{fig:4}), the values of the elastic energy per unit surface
area for all the bundles were calculated and are plotted as a
function of pressure in Fig. 5. It is clear that the elastic
energy, as expected, follows the structural transition shown in
Fig. \ref{fig:2}. The insets show that the relative increase in
elastic energy during collapse increases linearly with radius for
both SWNT and DWNT, which can be understood by the following
argument. The elastic energy per unit length of a nanotube of
radius $R$ before collapse is $U_{bc}=D\pi/R$. After collapse, the
tube has flat regions, which have no elastic energy, and bent
regions. The shape of the bent regions is invariant with respect
to change in radius $R$. That is, any increase in $R$ simply
increases the flat regions. Let $R'$ be an effective radius of the
bent regions. Then, the elastic energy of the collapsed nanotube
is $U_{ac}=D\pi/R'$. Since $R'$ is a fixed number that is
independent of $R$ (but less than $R$ in magnitude), the relative
increase in elastic energy on collapse is
\begin{equation}
\frac{U_{ac}-U_{bc}}{U_{bc}}=\left(\frac{D\pi/R'-D\pi/R}{D\pi/R}\right)=\frac{R}{R'}-1>0.
\end{equation}
This shows that the percent change in elastic energy on collapse
depends linearly on the radius of the nanotubes.

\section{Scaling of the response with radius, adhesion energy, and
bending elasticity.}

The main results so far, $1/R^3$ dependence of $p_c$ and that the
$p_c$ for $DWNT$ is a sum of the $p_c$ for packing of separate
SWNT bundles, can be derived from a formulation based on elastica
theory \cite{timoshenko,Frisch-Fay1962}. Assume that the response
of nanotube bundles to external loading can be calculated by
minimizing a properly defined potential energy functional
\cite{Tang_2cyl,Tang2005}. For a nanotube bundle of length $L$,
assume that deformations are primarily two-dimensional in the
cross section of the bundle.  Let $s$ denote length along a path
that traverses the graphene sheets in this 2D cross section. Let
$S$ represent the complete path and $S_P$ the part of this path on
which we apply external pressure. Tubes interact with each other
via an interatomic potential that has an attractive van der Waals
component and a short-range repulsion.  The latter component of
the interaction effectively prohibits two surfaces from
approaching each other too closely. Therefore, the surface $S$ can
be considered as consisting of two parts. Over the part of $S$
where the short-range repulsion creates a flat interface, $S_a$,
we say they are in contact; over the remaining surface, $S-S_a$,
we say that they are not in contact. The utility of this
partitioning is that a material property, the work of adhesion of
two graphene sheets, can be associated with $S_a$. The potential
energy per unit length of the bundle can be written as
\cite{Tang_2cyl,Tang2005}
\begin{equation}
\frac{V}{L}=\int_S \frac{D}{2}k^2  ds - W_a S_a -  \int_{S-S_a}
u_{vdw}^{ext} ds - \int_{S_T}{\bf T}\cdot{\bf v} ds +\int_S u_g
ds, \label{energy_1}
\end{equation}
where the first term represents energy due to bending of the
graphene sheet to mean curvature $k$, the second and third terms
capture adhesive van der Waals interactions, the fourth term
represents the work of external forces with $\bf T$, the external
traction, and $\bf v$, the displacement on the surface where
tractions are applied, and the last term is the energy of
formation of a flat sheet ($u_g$ is the energy of formation per
unit surface area of a graphene sheet). The attractive van der
Waals energy has been written in two parts. The first, $W_a S_a$,
captures regions where graphene sheets are in contact; $W_a$ is
the work of adhesion per unit area of bringing two \textit{flat}
nanotube walls from infinity to equilibrium separation.  For DWNT
and MWNT, there is a contribution to the work of adhesion due to
interlayer contact. Because layers deform together, this
contribution does not change with deformation.  It will therefore
vanish in a variation and for DWNT/MWNT $S_a$ can be identified as
the area of contact between outermost layers. The term
$u_{vdw}^{ext}$ represents interactions outside the contacting
regions. Once the tubes are in contact, the change in this term
with further deformation can be neglected
\cite{Tang_2cyl,Tang2005}. The scaling of the solution can be
extracted simply by a suitable normalization. Normalizing all
length scales in Eq. \ref{energy_1} by the radius, and dropping
terms that vanish in a variation, we obtain an expression for
potential energy per unit length, $v$,
\begin{equation}\label{beta}
v=\frac{RV}{DL}=\int_{\bar{S}} \frac{1}{2}{\bar{k}}^2 d\bar{s} -
\alpha \bar{S}_a +\int_{\bar{S}_T} {\bf \bar{b}}\cdot{\bf \bar{v}}
d\bar{s};\; \alpha=\frac{W_aR^2}{D};\;\mathbf{\bar{b}}=
\frac{\mathbf{T}R^3}{D}=\beta\mathbf{n};\;\beta =\frac{PR^3}{D}
\end{equation}
where ${\bf \bar{b}}$ is a dimensionless applied traction field,
$\beta$ is its (scalar) value for the case of fixed applied
pressure, and $\bf n$ is the unit normal. Our dimensionless
formulation implicitly assumes that no other length scale enters
into the problem, for example, through boundary conditions.  A
possible exception is the interlayer spacing.  Prior to collapse,
it has been shown that the deformation of two nanotubes in contact
is independent of this parameter\cite{Tang_2cyl,Tang2005}. In the
collapsed state the interlayer spacing perturbs only slightly the
solution obtained by neglecting it\cite{Tang_2cyl,Tang2005}.  On
this basis, we neglect its influence on our formulation; this
assumption is justified by good agreement between the predicted
scaling and simulation results.

Therefore, within our assumptions, the deformation depends only on
two dimensionless parameters, $\alpha$ and $\beta$.  If the
dominant influence on deformation is external pressure, then
events such as collapse or phase transitions will occur at
critical values of $\beta$, say at $\beta_c$.  This establishes
the scaling of critical pressure to be $P_c=\beta_c D/R^3$.  In
the absence of external pressure such events will occur at
critical values of $\alpha$, as already established
\cite{Tang_2cyl,Tang2005}.

For
commensurately packed MWNT, where difference in radius equals the
equilibrium separation between graphene sheets, this argument can
be extended to DWNT and MWNT.  First, we recognize that the shape of
any one shell in a deformed MWNT can be obtained from the shape
of another shell simply by a change of scale.  To build up
a deformed MWNT packing, we therefore first start with
a SWNT packing, say the shell with the largest diameter.
Consider Fig. 1 (a), a SWNT bundle,
in the case of deformations dominated by external pressure. At any
stage of the deformation the solution $\bar{k}$ is a function of
$\alpha,\beta$. Denote by ${\bf T}$ the surface tractions needed
to support the shape of this shell.  Now make an identical
copy of the deformed
bundle and reduce its diameter by a change in scale to a value just
small enough to fit inside the first shell. The
new shape is also a solution if we scale all the tractions
according to ${\bf{T}}=\beta D/R^3$.  Take the smaller
bundle, separate the nanotubes, and insert each into its
corresponding tube in the larger bundle. If we assume that the
interface between the walls of a DWNT cannot carry any shear
tractions, the resulting bundle is also a solution.  We note that
the tractions on the inner tube have to be provided by the outer
tube.  This leads to the conclusion that the net external
tractions we need to apply to the outer shell is
the sum of tractions needed to bring the two constituent SWNT's
to similar shapes. This is
easily generalized to a MWNT, establishing the fact that for
deformation to the same normalized shape, the needed applied
pressure is $p=\sum_{i=1}^n p_i$. All the nanotubes will collapse
simultaneously and so
\begin{equation}
p_c=\sum_{i=1}^n p_{ci}=\beta_c nD
\frac{1}{n}\sum_i^n\frac{1}{R_i^3}
\end{equation}
thus establishing the result that the collapse pressure for a MWNT
packing is the sum of collapse pressures of the constituent SWNT
packings, and providing the rationale for the effective radius
defined earlier.

As an independent test of this model, we have plotted the collapse
pressure as a function of $R_{eff}$ in Fig. 3 (b); it fits a
$1/R_{eff}^3$ relationship well.  The fit yields a value for
$\beta_c D_{eff}= 8.92$ eV.  If we define the surface of DWNT to
be a cylinder with radius $R_{eff}$, the energy per unit surface
area of DWNT is given by
\begin{eqnarray}\label{eq:EDfR}
u_e&=& \frac{1}{2\pi R_{eff}}\left(\frac{\pi D}{R_1}+
\frac{\pi D}{R_2}-W_a \pi(R_1+R_2)+2\pi u_g(R_1+R_2)\right)\nonumber\\
&=&\frac{D_{eff}}{2\tilde{R}^2}-W_a\frac{R_1+R_2}{2R_{eff}}+
u_g\frac{R_1+R_2}{R_{eff}},
\end{eqnarray}
where $1/\tilde{R}^2=({1/ 2R_1R_{eff}}+{1/ 2R_2R_{eff}})$, $R_1$
is the radius of the inner tube, and $R_2$ is the radius of the
outer tube. In Eq. \ref{eq:EDfR}, the first term corresponds to
the elastic energies of the inner and outer tubes, the second term
to the interaction energy of the two tubes and the third term to
the energy of formation of the two tubes. Eq. \ref{eq:EDfR} shows
that the energy per unit surface area of DWNT scales with the
inverse square of the length $\tilde{R}$. This quantity is readily
computed for different DWNT and Fig. 4(b) plots the single point
energy per unit surface area of five DWNT as a function of
$1/\tilde{R}^2$. A fit using \cite{Tang_2cyl,Tang2005} $u_g=0.765$
and $W_a=0.4$ yields a value of $D_{eff}=6.4$ eV, which is close
to twice $D_{SWNT}$. Together with the fit to collapse pressure,
the $DWNT$ data yield a value of $\beta_c^{DWNT}=1.39$, very close
in value to the that obtained from $SWNT$ simulations
($\beta_{c}^{SWNT}=1.31$).

\section{Summary}

To summarize, we use classical MD simulations to show that DWNT
bundles collapse at a critical pressure $p_{c}$ that, like in the
case of SWNT, varies as $1/R_{eff}^{3}$, where $R_{eff}$ is a
suitably defined effective radius. We find that the SWNT and DWNT
bundles show a $\sim$ 30\% hysteresis and  that the hexagonally
close packed lattice is completely restored in all SWNT and DWNT
bundles on decompression. Interestingly, we find that the $p_{c}$
of a DWNT bundle varies as the sum of the $p_{c}$ of the inner and
the outer tubes considered separately as SWNT bundles (a result we
derive analytically), demonstrating that the inner tube supports
the outer tube and that $D_{DWNT}{\sim}2D_{SWNT}$, where $D$ is a
bending stiffness.

\begin{center}
\textbf{ACKNOWLEDGEMENTS}
\end{center}

AKS thanks the Department of Science and Technology, Govt. of
India, for financial support. AJ's initial contribution to this
work was made during a stay at the Indian Institute of Science in
2005 as DuPont Chair. He would like to acknowledge the DuPont
company for their support. VG thanks the Centre for High Energy
Physics, Indian Institute of Science, for computational facilities
and Dr. Sachindeo Vaidya and Dr. Tarun Deep Saini for useful
discussions.

\textit{Note.} After the completion of our analysis, Ye et al.
\cite{Ye2005} published constant pressure MD simulations
demonstrating a hydrostatic pressure-induced structural transition
for \emph{isolated} DWNT. The values of the critical pressures
they obtain for isolated DWNT are 0.4 to 0.5 times  the values we
find for the same diameter DWNT arranged in a bundle.

\begin{center}
\textbf{APPENDIX: Calculation of mean curvatures for SWNT and DWNT
bundles}
\end{center}

The following algorithm was used for the calculation of the mean
curvature for a bundle of SWNT or DWNT. Each of the 16 tubes in
the system is an armchair tube $(n,n)$ with ten unit cells. It can
be shown that the total number of atoms per tube is $40n$. The
local curvature is calculated at every atom that belongs to the
middle eight unit cells ($32n$ atoms). The atoms belonging to the
unit cells at the ends of the tubes are not considered because
these atoms do not have the sufficient number of neighbors
required for our calculations (as will be clear later).

Each of the $32n$ atoms is considered one at a time. For each
atom, the coordinates of its three nearest neighbors and six next
nearest neighbors are found using a search algorithm. The central
atom's three nearest neighbors are used to define a plane passing
through them and the normal to this plane is found. This is
defined to be the new z-axis. The new x- and y-axes are suitably
defined to be mutually perpendicular.

A rotation matrix is now constructed using the components of the
normal. The matrix is then used to transform the coordinates of
the ten atoms (the central atom and its nine neighbors) to the new
coordinate system. In the new coordinate system, a quadratic
surface of the form $z=g(x,y)$ is fit to the ten points as
follows. The expanded form of the equation is given by
$z=ax^{2}+by^{2}+cxy+dx+ey+f$. This equation can be written
treating $(a,b,c,d,e,f)$ as the unknowns and
$(x^{2},y^{2},xy,x,y,z)$ as the coefficients. The coordinates of
the ten points give us ten equations in six unknowns. In matrix
notation, we have $[N]_{10\times6}[A]_{6\times1}=[Z]_{10\times1}$,
where $[A]$ is the matrix to be determined. The values of
$(a,b,c,d,e,f)$ are obtained by calculating $[A]$ using the
relation $[A]=([N]^{T}[N])^{-1}[N]^{T}[Z]$.

The mean curvature of a surface, as defined above, is given by
\cite{Dubrovin1992}
\begin{equation}
H=\nabla\cdot\left(\frac{\nabla{g}}{\sqrt{1+|\nabla{g}|^{2}}}\right).
\end{equation}
The value of $H$ at the central atom is now calculated using the
values of $(a,b,c,d,e,f)$ and the coordinates of the central atom.

This process is repeated for all the $32n$ atoms of the tube. The
local curvature values at atoms of the other fifteen tubes in the
bundle are similarly calculated to yield a total of $512n$ values.
The average curvature for the bundle is simply the mean of these
$512n$ values. For bundles of DWNT, the same procedure is used,
treating the inner and the outer tubes as separate SWNT and
averaging over atoms in 32 tubes.

This method gave good results for all tubes except the very small
(5,5) tube, which cannot be well approximated by a smooth cylinder
even at 0 K. The calculated mean curvature, using the method
described above, for an optimized (5,5) tube at 0 K differs from
the value of the curvature of a cylinder of the same radius (given
by $1/radius$) by more than 5 \%.

\clearpage

\begin{table}
\begin{tabular}{lllllll}
  \hline
  $E_{bond}(R)=\frac{1}{2}K_{b}(R-R_{0})^{2}$ & $R_{0}$ & 1.39 $\AA$ & $K_{b}$ & 1050 (kcal/mol)/$\AA^{2}$\\
  $E_{angle}(\theta)=\frac{1}{2}K_{\theta}(\cos\theta-\cos\theta_{0})^2$ & $\theta_{0}$ & 120$^{\circ}$ & $K_{\theta}$ &
  100 (kcal/mol)/rad$^{2}$\\
  $E_{torsion}(\phi)=\frac{1}{2}V\{1-\cos[n(\phi-\phi_{0})]\}$ & $\phi_{0}$ & 180$^{\circ}$ & $V$ &
  25.0 kcal/mol & $n$ & 2\\
  $E_{inv}(\Psi)=\frac{1}{2}\frac{K_{i}}{(\sin\Psi_{0})^2}(\cos\Psi-\cos\Psi_{0})^2$ & $\Psi_{0}$ & 0$^{\circ}$ & $K_{i}$ &
  40 (kcal/mol)/rad$^{2}$\\
  $E_{vdW}(R)=D_{0}\left\{\left(\frac{R_{0}}{R}\right)^{12}-2\left(\frac{R_{0}}{R}\right)^{6}\right\}$ & $R_{0}$ & 3.8983 $\AA$ & $D_{0}$ & 0.0951 kcal/mol\\
  \hline
\end{tabular}
\caption{Parameters for the C\_R atom type (sp$^2$ hybridized
carbon atom involved in resonance), in DREIDING \cite{Mayo1990}, a
standard generic macromolecular force field used in all our
molecular dynamics simulations.}\label{Table1}
\newpage
\end{table}

\begin{figure}[tbp]
\begin{center}
\leavevmode
\includegraphics[width=\textwidth]{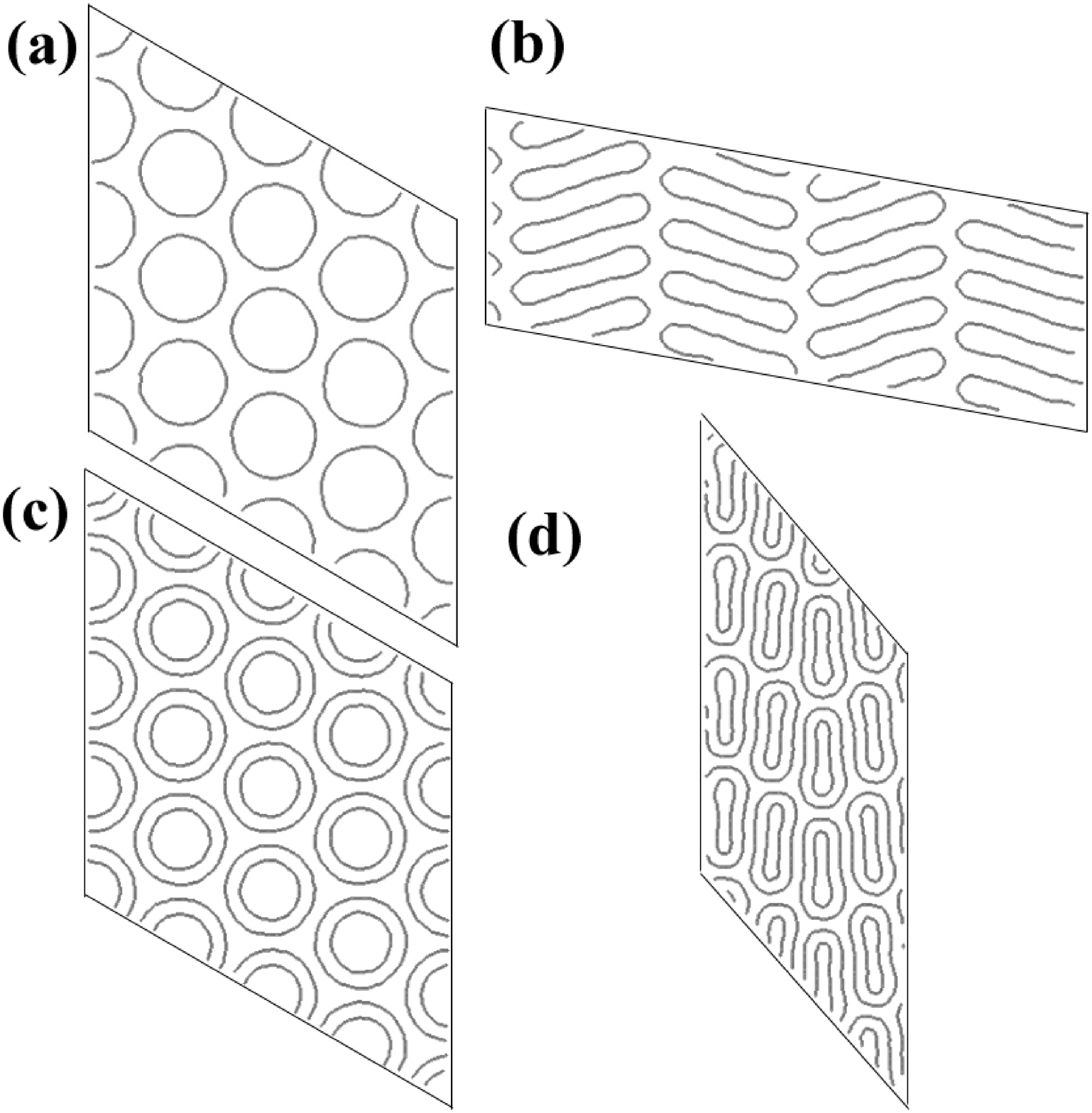}
\caption{The upper figures show a 4 $\times$ 4 bundle of (10,10)
SWNT at (a) $p = 1.0$ Atm (before collapse) and (b) $p = 6.0$ GPa
(after collapse). The bottom figures show a 4 $\times$ 4 bundle of
(10,10)@(15,15) DWNT at (c) $p = 1$ Atm (before collapse) and (d)
$p = 10.2$ Gpa (after collapse).} \label{fig:1}
\end{center}
\end{figure}

\clearpage

\begin{figure}[tbp]
\begin{center}
\leavevmode
\includegraphics[width=\textwidth]{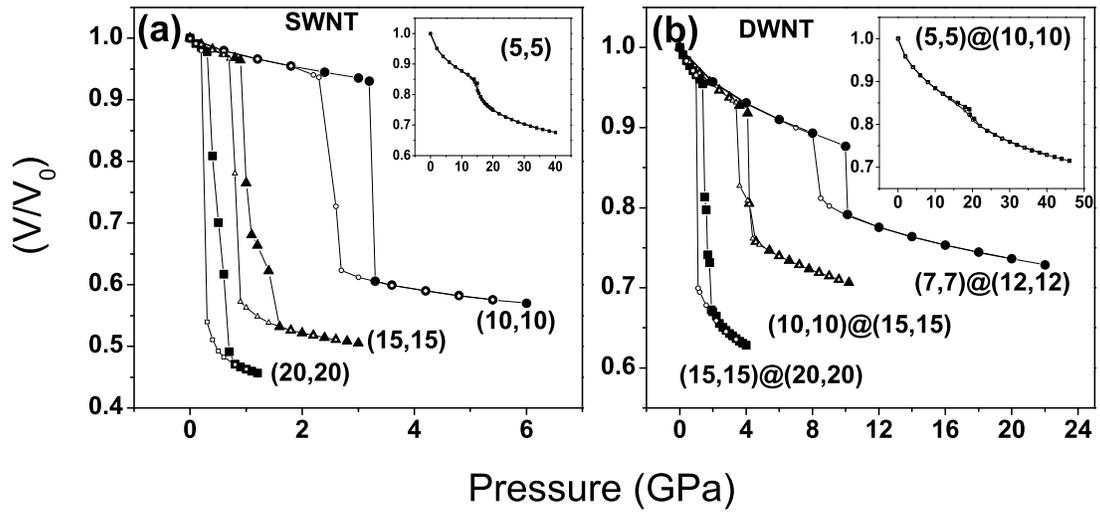}
\caption{Reduced volume
($V/V_{0}$) as a function of applied hydrostatic pressure for (a)
SWNT and (b) DWNT bundles. The loading (solid symbols) and
unloading (open symbols) curves clearly show hysteresis.}
\label{fig:2}
\end{center}
\end{figure}

\clearpage

\begin{figure}[tbp]
\begin{center}
\leavevmode
\includegraphics[width=\textwidth]{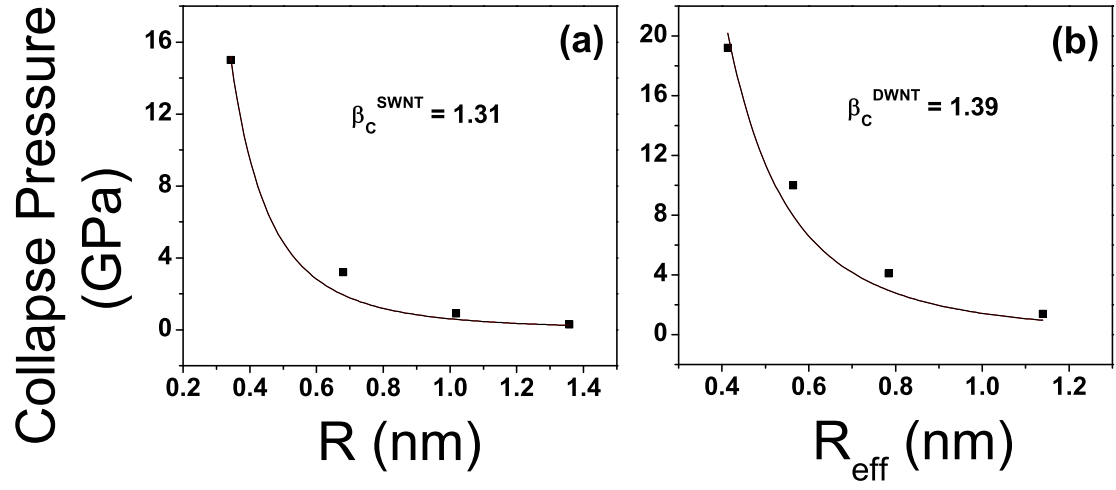}
\caption{Critical collapse pressure ($p_{c}$) as a function of (a)
SWNT radius and (b) DWNT effective radius defined in the text. We
estimate the values of $\beta_{c}$ for both SWNT and DWNT using
Eq. \ref{beta}.} \label{fig:3}
\end{center}
\end{figure}

\clearpage

\begin{figure}[tbp]
\begin{center}
\leavevmode
\includegraphics[width=\textwidth]{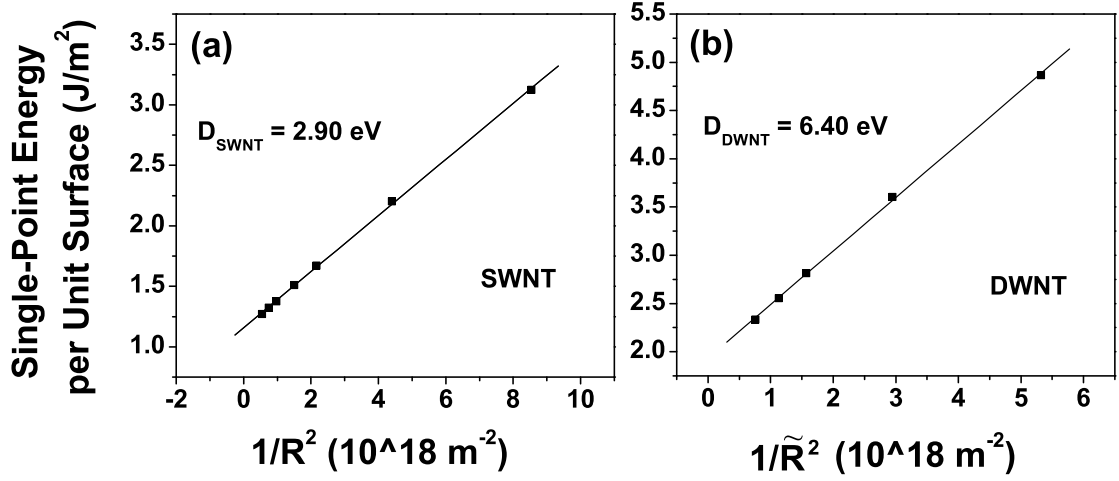}
\caption{Single point energy per unit surface area at 0 K as a
function of $1/R^{2}$ for seven isolated SWNT (a) and as a
function of $1/\tilde{R}^{2}$ for five isolated DWNT (b). The
value of $D$, the bending stiffness, is obtained from the fit to
Eq. \ref{eq:EDfR}.} \label{fig:4}
\end{center}
\end{figure}

\clearpage

\begin{figure}[tbp]
\begin{center}
\leavevmode
\includegraphics[width=\textwidth]{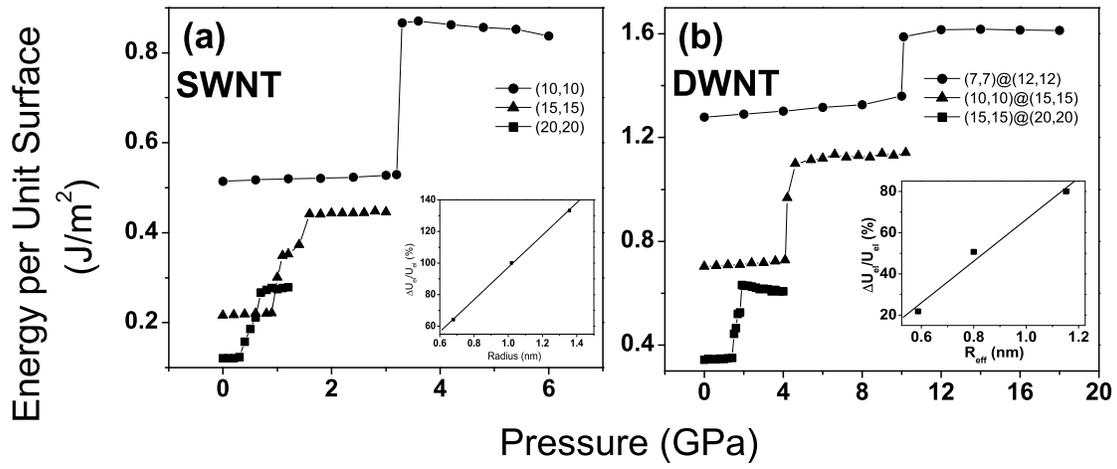}
\caption{Energy per unit surface area as a function of applied
hydrostatic pressure for (a) SWNT, and (b) DWNT bundles. Notice
the correspondence with Fig. \ref{fig:2}. The insets show the
relative increase in elastic energy during collapse as a function
of radius.} \label{fig:5}
\end{center}
\end{figure}

\end{document}